\documentclass[aps,prl,twocolumn,showpacs,superscriptaddress,floatfix]{revtex4-1}

\usepackage{graphicx}
\usepackage{dcolumn}
\usepackage{bm}
\usepackage{amssymb}
\usepackage{microtype}
\usepackage{xfrac}

\begin{document}

\title{Statics and Dynamics of the Highly Correlated Spin Ice, Ho$_2$Ge$_2$O$_7$}

\author{A.~M.~Hallas}
\affiliation{Department of Chemistry, University of Manitoba, Winnipeg, MB, R3T 2N2, Canada}

\author{J.~A.~M. Paddison}
\affiliation{Department of Chemistry, University of Oxford, Inorganic Chemistry Laboratory, South Parks Road, Oxford, OX1 3QR, United Kingdom}
\affiliation{ISIS Facility, Rutherford Appleton Laboratory, Chilton, Didcot, OX11 0QX, United Kingdom}

\author{H.~J.~Silverstein}
\affiliation{Department of Chemistry, University of Manitoba, Winnipeg, MB, R3T 2N2, Canada}

\author{A.~L.~Goodwin}
\affiliation{Department of Chemistry, University of Oxford, Inorganic Chemistry Laboratory, South Parks Road, Oxford, OX1 3QR, United Kingdom}

\author{J.~R.~Stewart}
\affiliation{ISIS Facility, Rutherford Appleton Laboratory, Chilton, Didcot, OX11 0QX, United Kingdom}

\author{J.~G.~Cheng}
\affiliation{Texas Materials Institute, University of Texas at Austin, Austin, TX, 78712, USA}

\author{J.~S.~Zhou}
\affiliation{Texas Materials Institute, University of Texas at Austin, Austin, TX, 78712, USA}

\author{J.~B.~Goodenough}
\affiliation{Texas Materials Institute, University of Texas at Austin, Austin, TX, 78712, USA}

\author{E.~S.~Choi}
\affiliation{National High Magnetic Field Laboratory, Florida State University, Tallahassee, FL, 32306-4005, USA}

\author{G.~Ehlers}
\affiliation{Quantum Condensed Matter Division, Oak Ridge National Laboratory, Oak Ridge, TN, 37831-6475, USA}

\author{J.~S.~Gardner}
\affiliation{NIST Center for Neutron Research, Gaithersburg, MD, 20899-6102, USA}
\affiliation{Indiana University, 2401 Milo B. Sampson Lane, Bloomington, IN, 47408, USA}

\author{C.~R.~Wiebe}
\affiliation{Department of Chemistry, University of Manitoba, Winnipeg, MB, R3T 2N2, Canada}
\affiliation{National High Magnetic Field Laboratory, Florida State University, Tallahassee, FL, 32306-4005, USA}
\affiliation{Department of Chemistry, University of Winnipeg, Winnipeg, MB, R3B 2E9 Canada}

\author{H.~D.~Zhou}
\affiliation{National High Magnetic Field Laboratory, Florida State University, Tallahassee, FL, 32306-4005, USA}
\affiliation{Department of Physics and Astronomy, University of Tennessee, Knoxville, TN, 37996-1200, USA}

\begin{abstract}
The pyrochlore Ho$_2$Ge$_2$O$_7$ is a new highly correlated spin ice material. Physical property measurements including x-ray diffraction, dc susceptibility and ac susceptibility, confirm that it shares the distinctive characteristics of other known spin ices. Polarized neutron scattering measurements on a powder sample, combined with reverse Monte Carlo (RMC) refinements, give unique information about the spin ice state in Ho$_2$Ge$_2$O$_7$. RMC refinements are used to fit the powder magnetic diffuse scattering and predict the single crystal magnetic scattering of Ho$_2$Ge$_2$O$_7$, demonstrating consistency with spin ice behavior.
\end{abstract}


\maketitle
The pyrochlore lattice has provided condensed matter physicists with a wide variety of magnetic behaviors to study, due to the presence of geometric frustration \cite{GGG}. Pyrochlores, with formula A$_{2}$B$_{2}$O$_{7}$, are composed of two interpenetrating sublattices of corner-sharing tetrahedra where the A and B sites each form the vertices of one such network. The A-site is commonly occupied by a magnetic rare earth ion, which allows for highly frustrated interactions. In some materials, the frustration can be overcome at sufficiently low temperatures and long range order can be achieved. However, in many magnetic pyrochlores the frustrated interactions are prohibitive to long range ordering, and instead novel ground states, such as spin liquids, spin glasses or spin ices, are adopted \cite{Greedan}.

The spin ice state was first observed in Ho$_2$Ti$_2$O$_7$ by Harris {\em et al.} in 1997 \cite{HoTi2}; since that time spin ices have been a subject of active experimentation, allowing theorists to come a long way towards understanding this remarkable ground state. In spin ices, the local crystal field acting on the magnetic ions results in nearly perfect Ising spins which align along the axis that joins the centers of two neighboring tetrahedra. When combined with overall ferromagnetic nearest-neighbour interactions, this results in short-range magnetic order with two spins pointing inwards and two spins pointing outwards from the center of each tetrahedron.

Despite significant interest in this class of compounds, only a handful of spin ice materials have been discovered to date, including the titanates, A$_2$Ti$_2$O$_7$ \cite{HoTi1, HoTi2, DyTi1, DyTi2}, the stannates, A$_2$Sn$_2$O$_7$ \cite{HoSn1, DySn1}, and more recently, the germanates, A$_2$Ge$_2$O$_7$ \cite{DyGeO, HoGeO1, HoGeO2} (A~=~Ho,~Dy). Spin ices have recently garnered further attention as the hosts of quasiparticles resembling magnetic monopoles \cite{Castelnovo, monopole1, monopole2}. This paper details the synthesis and characterization of a new highly correlated spin ice candidate, Ho$_2$Ge$_2$O$_7$. Reverse Monte Carlo (RMC) refinement is utilized to confirm that the ice rules are obeyed in this material and to demonstrate the suitability of this material to act as a monopole host.

\begin{figure}[tbp]
\linespread{1}
\par
\includegraphics[width=3.3in]{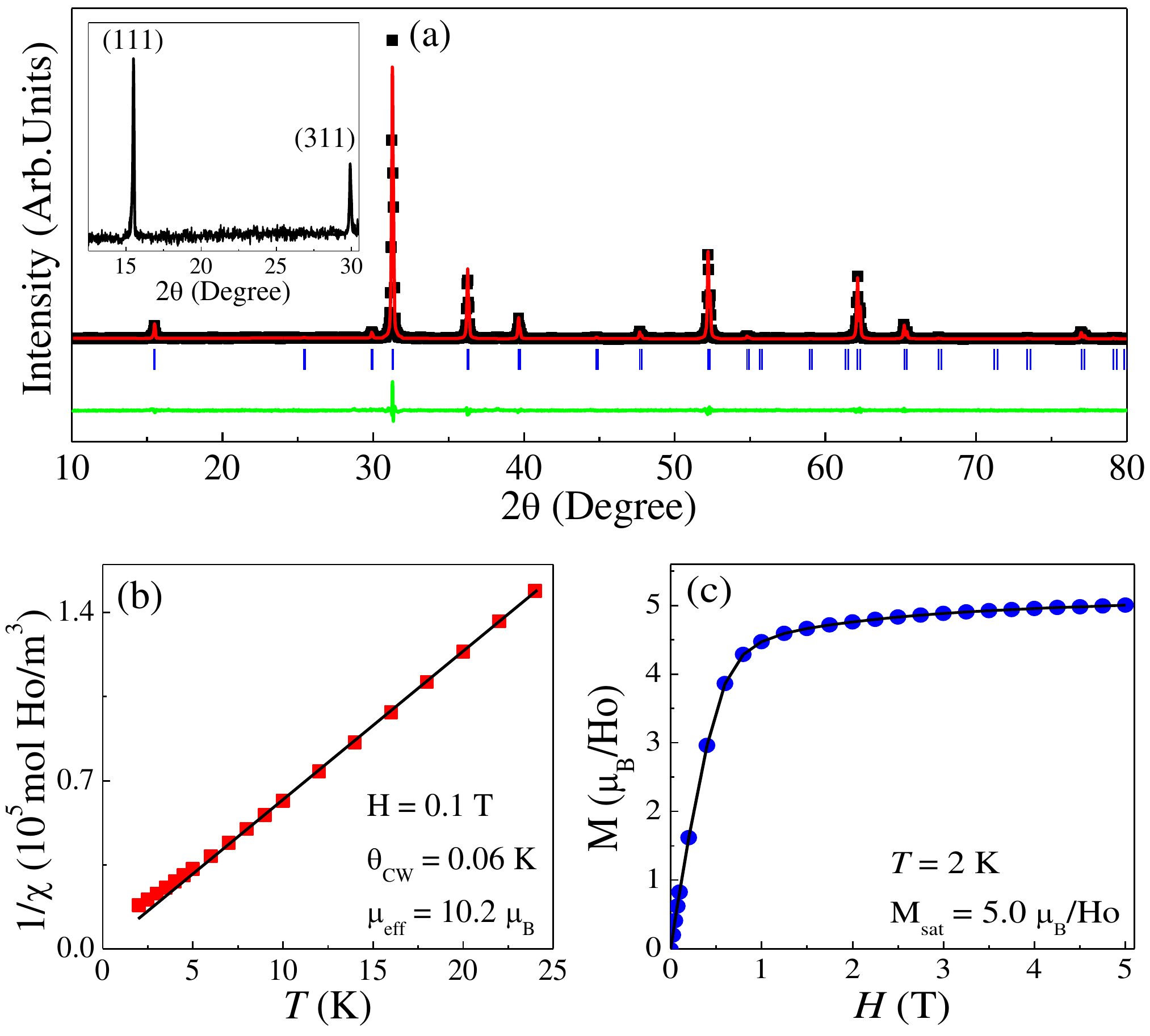}
\par
\caption{(a) Rietveld refinement for Ho$_2$Ge$_2$O$_7$ (Cu K$_{\alpha}$, $\lambda$~=~1.5418 {\AA}). Squares are the measured data, the solid (red) curve is the calculated fit, the vertical marks indicate the Bragg peak positions, and the bottom (green) curve is the difference between measured and calculated intensity. Insert: low 2$\theta$ region containing the pyrochlore superlattice peaks. (b) The temperature dependence of the dc susceptibility for Ho$_2$Ge$_2$O$_7$ measured with $H$ = 0.1 T. (c) The dc magnetization for Ho$_2$Ge$_2$O$_7$ measured at $T$~=~2~K.}
\end{figure}

When prepared under ambient pressure with a conventional solid state reaction, Ho$_2$Ge$_2$O$_7$ is a pyrogermanate with a tetragonal structure \cite{Cava}. Synthesis of polycrystalline Ho$_2$Ge$_2$O$_7$ in the pyrochlore phase was performed with a high-temperature and high-pressure technique. Stoichiometric amounts of Ho$_2$O$_3$ and GeO$_2$, wrapped in gold foil, were compressed to 7 GPa and heated to 1000~$^\circ$C in a Walker-type, multi-anvil press. The quality of the sample was assessed using a Guinier image plate x-ray diffractometer with a rotating copper anode source. Rietveld refinement with FullProf \cite{FullProf} showed no discernible impurities.

The x-ray diffraction pattern of Ho$_2$Ge$_2$O$_7$, shown in Fig. 1(a), contains superlattice peaks at 2$\theta$ $\approx$ 15$^{\circ}$ and 30$^{\circ}$ for the (111) and (311) Bragg peaks respectively, indicating a pyrochlore type lattice (left insert of Fig. 1(a)). Rietveld refinement of the XRD pattern confirmed the face-centered cubic pyrochlore phase (Fd$\overline{3}$m, No. 227) and the absence of any tetragonal pyrogermanate impurity. The room temperature lattice parameter was determined
to be $a$ = 9.9026(6) {\AA}, which agrees well with the reported value \cite{HoGeO2}. The lattice parameter is significantly reduced from previously synthesized holmium pyrochlores by the introduction of the much smaller germanium cation onto the B-site (Ho$_2$Sn$_2$O$_7$ and Ho$_2$Ti$_2$O$_7$ have $a$ = 10.3762 {\AA} and $a$ = 10.1059 {\AA} respectively \cite{HoTiAC}).

The dc magnetic susceptibility, $\chi$, of Ho$_2$Ge$_2$O$_7$ measured from 1.8~K to 20~K with a field intensity of 0.1~T shows no sign of long range magnetic ordering (Fig. 1(b)). The Curie-Weiss temperature for this material is $\theta_{\text {CW}}$~=~0.06~K. Small Curie-Weiss temperatures arise in spin ice materials due to the comparable scales of the magnetic exchange, $J_{\text {nn}}$, and ferromagnetic dipolar interactions, $D_{\text {nn}}$, between nearest neighbors \cite{HoGeO1}. The Curie-Weiss temperatures for Ho$_2$Sn$_2$O$_7$ and Ho$_2$Ti$_2$O$_7$ are 1.8~K and 1.9~K respectively. The much smaller $\theta_{\text {CW}}$ in Ho$_2$Ge$_2$O$_7$ results from the reduction in lattice parameter \cite{HoGeO1}. The effective magnetic moment in Ho$_2$Ge$_2$O$_7$, derived from the dc susceptibility, is 10.2(1)~$\mu_{\text B}$ per holmium. However, the magnetization, measured at 2 K (Fig. 1(c)), saturates at $\sim$5 $\mu_{\text B}$ per holmium. The saturated moment in polycrystalline spin ice samples is reduced to half the value of the magnetic moment at each site due to the $<$111$>$ local Ising magnetic anisotropy and powder averaging.

\begin{figure}[tbp]
\linespread{1}
\par
\includegraphics[width=3.2in]{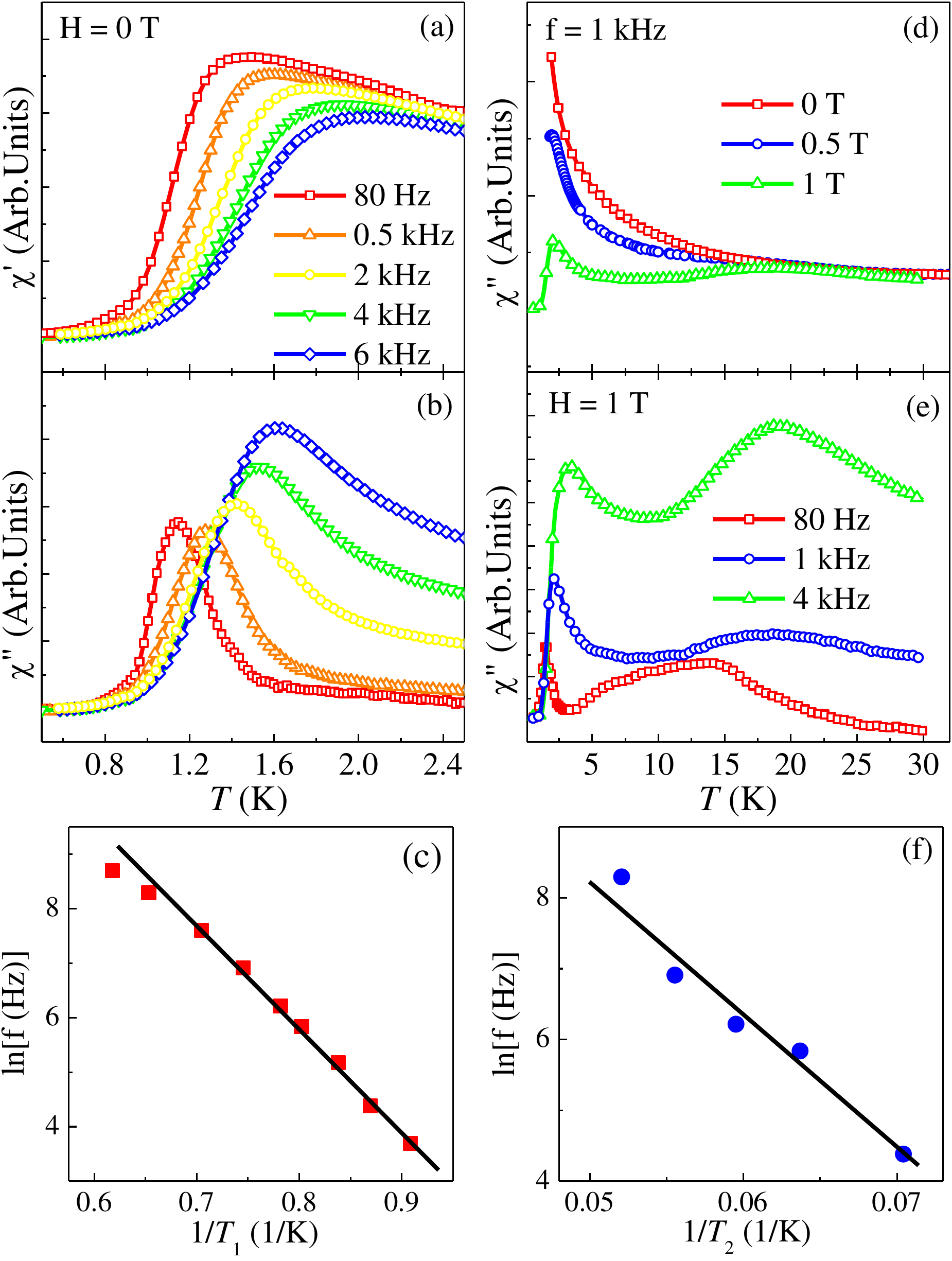}
\par
\caption{Temperature dependence of (a) the real part, $\chi^{\prime}$, and (b) the imaginary part $\chi^{\prime\prime}$ of the ac susceptibility for Ho$_2$Ge$_2$O$_7$ below 3 K; (c) Arrhenius law fit of the low temperature peak position in $\chi^{\prime\prime}$; (d) High temperature $\chi^{\prime\prime}$ at $f$ = 1 kHz with varying field strength; (e) High temperature $\chi^{\prime\prime}$ at $\mu_{\text 0}$H = 1 T with varying frequency; (f) Arrhenius law fit of the high temperature peak position in $\chi^{\prime\prime}$}
\end{figure}

The ac susceptibility measurements on Ho$_2$Ge$_2$O$_7$ were made using frequencies ranging from 80 Hz to 6 kHz in dc fields of 0~T to 1~T with temperatures between 0.5~K and 30~K. The real part of the ac susceptibility, $\chi^{\prime}$, below 3~K is shown in Fig.~2(a). With an applied frequency of 80~Hz, the susceptibility drops at 1.3~K and approaches zero in the limit of 0.5~K. Concurrently, the imaginary part, $\chi^{\prime\prime}$, (Fig.~2(b)) contains a single maximum at $\sim$1.2~K. In both the real and imaginary parts, the peak position shifts towards higher temperature with increasing frequency. The peak position in $\chi^{\prime\prime}$, $T_1$, is frequency dependent and can be fit to an Arrhenius law, $f=f_{0}\exp(-E_{1}/(k_{B}T_{1}))$, which yields $E_{1}/k$~=~20(2)~K (Fig.~2(c)). At high temperatures with a frequency of 1~kHz a second peak begins to emerge in $\chi^{\prime\prime}$ at $\sim$18~K with increasing field strength (Fig.~2(d)). Similarly to the first peak, the position of the peak when measured in a 1~T field is frequency dependent (Fig.~2(e)) and can likewise be fit to an Arrhenius law. The fit for the second peak, $T_{2}$, gives an energy barrier of $E_{2}/k$~=~196(10)~K (Fig.~2(f)).

The presence and position of these peaks in the imaginary part of the ac susceptibility for Ho$_2$Ge$_2$O$_7$ is consistent with observations for other known spin ices. The energy barrier of 20~K in the low temperature, spin freezing region is related to local spins adopting an ice-like state. The comparable energy barriers for Ho$_2$Sn$_2$O$_7$ and Ho$_2$Ti$_2$O$_7$ are 19.6~K and 27.5~K respectively \cite{HoTiAC}. The second, high temperature, energy barrier for Ho$_2$Ge$_2$O$_7$ of 196~K relates to a thermally activated region and the first crystal field excitation \cite{Ehlers}.

Neutron polarization analysis was performed on a 290~mg powder sample of Ho$_2$Ge$_2$O$_7$ using the D7 diffuse scattering spectrometer at the Institut Laue-Langevin \cite{D7}. The instrument was run in XYZ polarization analysis mode to allow the separation of the magnetic scattering from the nuclear-coherent, isotope-incoherent and spin-incoherent components. An incident wavelength of 4.855~{\AA} was selected, allowing $Q$ space coverage from~0.25~{\AA}$^{-1}$ to 2.5~{\AA}$^{-1}$. Measurements were taken at four temperatures, 50~mK, 1.3~K, 3.6~K, and 10.6~K using a dilution refrigerator and a $^4$He cryostat.

The magnetic diffuse scattering of Ho$_2$Ge$_2$O$_7$, shown in Fig. 3(a), has the characteristic shape for a spin ice at 50~mK. The minima observed at $Q$ $\approx$ 1.3~{\AA}$^{-1}$ and 2.2~{\AA}$^{-1}$ are likely the result of pinch-point singularities. It is the dipolar correlations in spin ices that give rise to pinch-points, a defining feature of these materials \cite{monopole2}. However, they can only be directly observed in a single crystal diffraction experiment with sufficient resolution. In Ho$_2$Ge$_2$O$_7$, the (002) and (222) pinch-points would give rise to a minima at $Q$ = 1.268 {\AA}$^{-1}$ and $Q$ = 2.196 {\AA}$^{-1}$ respectively, which agrees with the observed minima positions in Fig. 3(b). The (111) pinch-point would correspond to a minimum at $Q$ = 1.098 {\AA}$^{-1}$ which is not clearly defined as it overlaps with the (002) minimum. The characteristic minima observed in the magnetic diffuse scattering suggest that Ho$_2$Ge$_2$O$_7$ is a likely spin ice.

The temperature dependence of the magnetic diffuse scattering in spin ices has been previously measured by Mirebeau {\em et al.} for Ho$_2$Ti$_2$O$_7$ \cite{Mirebeau}. The magnetic diffuse scattering for Ho$_2$Ge$_2$O$_7$ was measured at four temperatures (Fig. 3(a)) to determine what, if any, effect the stronger correlations in this spin ice would have on the higher temperature scattering. In Ho$_2$Ge$_2$O$_7$, the differences in magnetic diffuse scattering between 50~mK and 1.3~K are quite minimal, indicating that the spin ice state is, essentially, fully formed by 1.3~K. At 3.6~K the ice-like scattering is still prominent with some broadening of features as a larger portion of spins are disordered at this temperature. Characteristic ice-like scattering is still visible even at 10.6~K; however, at this temperature the majority of spins are disordered. These effects of temperature on the magnetic diffuse scattering are consistent with the results for other spin ices.

\begin{figure}[tbp]
\linespread{1}
\par
\includegraphics[width=3.4in]{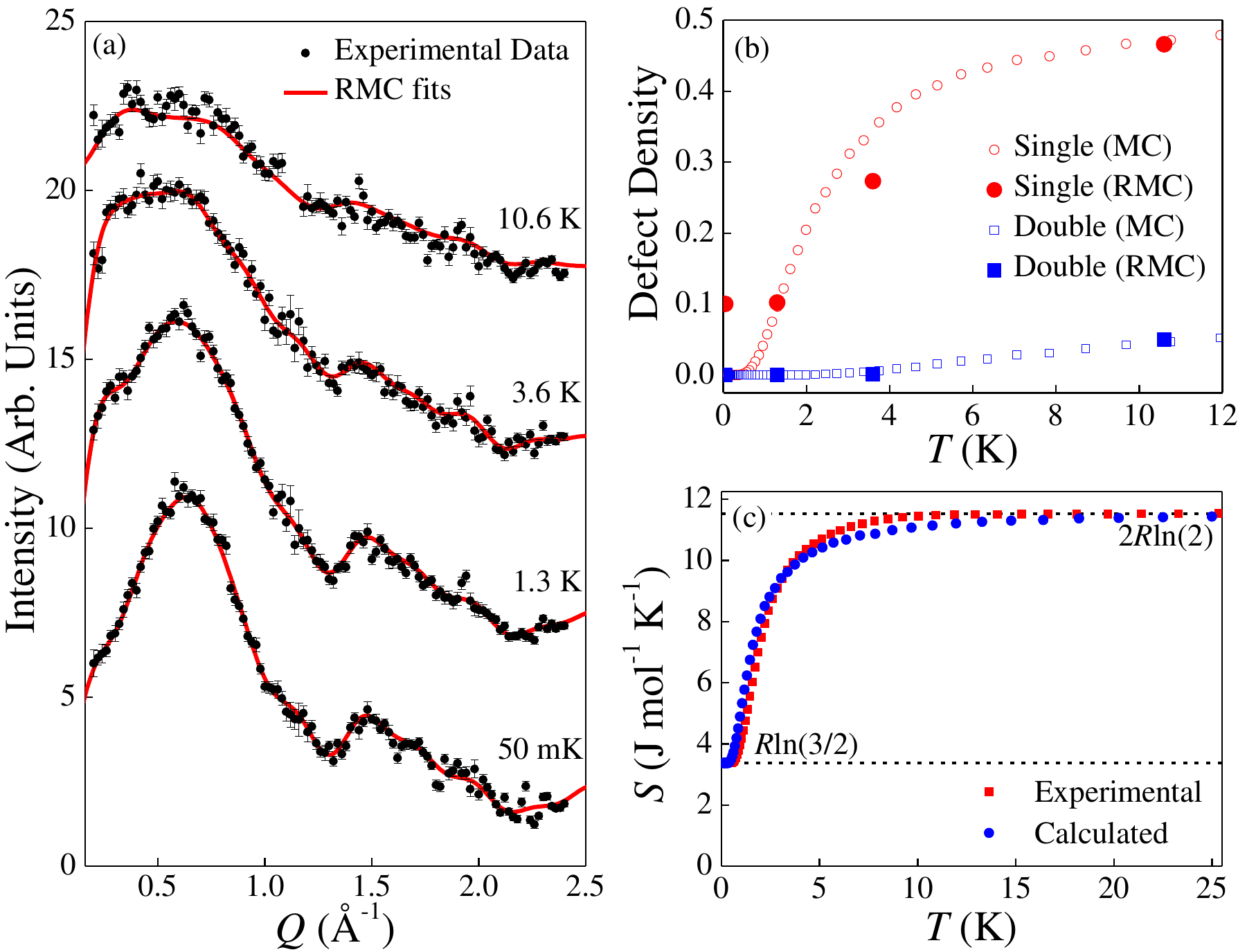}
\par
\caption{(a)~Temperature variation in the magnetic diffuse neutron scattering of Ho$_2$Ge$_2$O$_7$ with RMC fits. Error bars represent $\pm$1$\sigma$ propagated from the statistical uncertainty of the raw counts. (b)~Density of ice-rules defects as a function of temperature for Ho$_2$Ge$_2$O$_7$. (c)~Comparison of theoretical magnetic entropy for Ho$_2$Ge$_2$O$_7$ with experimental results.}
\end{figure}

The reverse Monte Carlo (RMC) technique was employed to fit the powder data and examine the temperature evolution of the magnetism in Ho$_2$Ge$_2$O$_7$ in more detail. This approach has been previously shown to produce accurate models of frustrated magnetic structures based on powder data \cite{JoePRL}. The RMC method does not consider a magnetic Hamiltonian, but instead involves refining the orientations of spins within a large configuration to obtain the best possible fit to data. The only constraint used in the following refinements is that the spins behave as purely Ising variables (i.e., $\mathbf{S}_{i}={S_{i}}\mathbf{z}_{i}$, with Ising pseudo-spin $S_{i}$~=~$\pm$1 directed along one of the cubic $\mathbf{z}_{i}$~=~$<$111$>$ axes). The powder scattering intensity is calculated from the RMC spin configurations using the exact expression for a magnetically anisotropic system \cite{Physics}. Refinements are performed using spin configurations of 5$^3$ crystallographic unit cells (2000 spins), using periodic boundary conditions, and calculated quantities are averaged over at least 16 independent configurations.

The RMC fits obtained for Ho$_2$Ge$_2$O$_7$ show excellent agreement with the experimental data at all four temperatures (Fig.~3(a)). The fraction of ice rules defects (defect density) which appear in the RMC spin configurations at each temperature can be calculated from these fits. These values represent an upper bound on the true defect density because RMC fitting is a stochastic approach which will tend to produce the most disordered spin configurations compatible with experiment. The overestimation is most serious below $\sim$1~K, where the anticipated defect density and the error have the same order of magnitude. Nevertheless, at higher temperatures, the ``experimental" RMC defect densities can be usefully compared to the theoretical values determined from a spin ice Hamiltonian. This provides an independent check on whether spin ice behavior describes the basic dynamics of Ho$_2$Ge$_2$O$_7$. Accordingly, in Fig.~3(b) the RMC defect densities are compared with the theoretical values, which are determined from direct Monte Carlo simulations of the nearest-neighbour spin ice model using an effective interaction $J_{\text {eff}}$~=~1.63~K \cite{HoGeO1}. As anticipated, the RMC upper bound at 50~mK of $\sim$10\% is anomalously large. There is, however, qualitative agreement with the experimental single defect densities at $T \geq 1.3$~K. The RMC density of double defects is zero at low temperatures and remains small at 10.6 K, in accordance with the theoretical prediction. This agreement with the nearest-neighbour spin ice model suggests ice-like dynamics are present in Ho$_2$Ge$_2$O$_7$.

\begin{figure}[tbp]
\linespread{1}
\par
\includegraphics[width=3.2in]{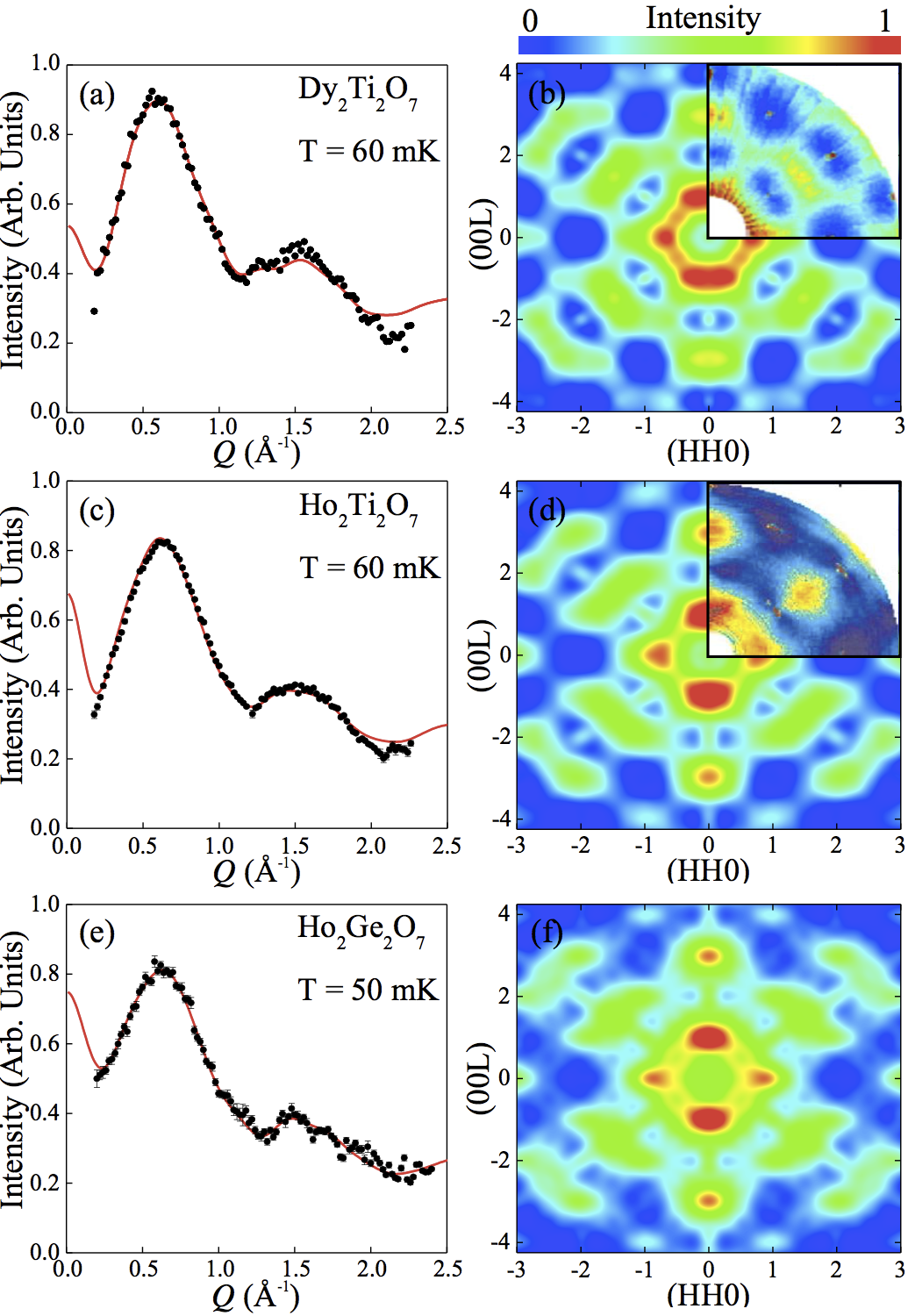}
\par
\caption{Left: Fits to experimental powder magnetic scattering data for the spin ice materials (a) Dy$_2$Ti$_2$O$_7$; (c) Ho$_2$Ti$_2$O$_7$; (e) Ho$_2$Ge$_2$O$_7$. Data points are shown as black circles and RMC fits as red lines. The fits shown are an average of 64 independent refinements. Error bars represent $\pm$1$\sigma$ propagated from the statistical uncertainty of the raw counts. Right: single crystal magnetic scattering predicted from RMC configurations for (b) Dy$_2$Ti$_2$O$_7$; (d) Ho$_2$Ti$_2$O$_7$; (f) Ho$_2$Ge$_2$O$_7$. Patterns are averaged over 64 configurations, with further averaging performed over regions of $3^{3}$ unit cells within each configuration \cite{Butler}. Experimental single crystal data for Dy$_2$Ti$_2$O$_7$ (From T. Fennell {\em et al.} \cite{DyTi2}) and Ho$_2$Ti$_2$O$_7$ (From T. Fennell {\em et al.} \cite{monopole2}) are shown as inserts for comparison.}
\end{figure}

The magnetic heat capacity for Ho$_2$Ge$_2$O$_7$ is shown in Zhou \emph{et al.} \cite{HoGeO1}. The integration of the magnetic heat capacity divided by temperature, $C_{\text {mag}}/T$, gives magnetic entropy $S_{\text {mag}} = 8.2~\rm{J mol}^{-1}_{\rm {tet}}\rm{K}^{-1}$ (Fig. 3(c)). The zero point entropy is calculated by taking the difference of this value and the expected value for an Ising system, $S = 2R\ln(2)$ per tetrahedron, which gives the characteristic spin ice zero point entropy, $S_{0} = R\ln(3/2)$. Using statistical mechanics and the quasi-particle picture developed by Castelnovo {\em et al.} \cite{Castelnovo,Castelnovo2}, an approximate expression for the magnetic entropy of a spin ice can be derived in terms of the number of ice rules defects:
\begin{eqnarray}
\frac{S}{R}&=&-x\ln\left (\frac{2x}{3}\right )-f_{1}\ln\left (\frac{f_{1}}{2}\right )-f_{2}\ln(2f_{2})\\
&&\text{where  } x = 1-f_{1}-f_{2}
\end{eqnarray}
where $f_{1}$ is the fraction of tetrahedra with a single ice rules defect and $f_{2}$ is the fraction of tetrahedra with a double defect. Spin configurations were generated using direct Monte Carlo simulations of the nearest-neighbor spin ice model, as described above. The entropy calculated from these spin configurations using Eq.~1 shows surprisingly good agreement with experiment (Fig. 3(c)), given the assumption of independent tetrahedra in Eq.~1 and the use of the simple nearest-neighbour spin ice model.

However, it is known that the nearest-neighbour spin ice model does not fully describe real spin ice materials. For a better description it is necessary to also consider the magnetic dipolar interaction \cite{model, HoTi1} and possible smaller exchange interactions beyond nearest-neighbors \cite{Joe16}. These interactions slightly favor certain ice rules configurations over others, and give rise to subtle differences between single crystal neutron scattering patterns for different spin ice materials. Obtaining a single crystal of Ho$_2$Ge$_2$O$_7$ in order to investigate such effects is currently unfeasible due to the high pressure required for synthesis. However, the RMC approach can be used to predict the single crystal diffuse scattering from powder data \cite{JoePRL}. By using the collective `loop' spin-flips described in \cite{Melko} as the basic RMC move, it is possible to enforce the ice rules. Thus, the nearest neighbor correlations are fixed and the sensitivity of the RMC refinement to small variations in further neighbor correlations can be maximized.

The effectiveness of this approach was first tested by fitting powder data collected on D7 for the two canonical spin ices Dy$_2$Ti$_2$O$_7$ and Ho$_2$Ti$_2$O$_7$ (Fig.~4(a) and (c)). The single crystal scattering was then calculated from the RMC spin configurations and compared with actual experimental data (Fig. 4(b) and (d)). The close agreement between experimental single crystal data and the RMC predictions demonstrates the effectiveness of the approach. The powder data for Ho$_2$Ge$_2$O$_7$ were then fitted in the same way with a good fit to data obtained (Fig.~4(e)). This result places a lower bound of zero on the density of ice rules defects present in Ho$_2$Ge$_2$O$_7$ at 50~mK. The predicted single crystal scattering for Ho$_2$Ge$_2$O$_7$ (Fig.~4(f)) differs slightly from both Ho$_2$Ti$_2$O$_7$ and Dy$_2$Ti$_2$O$_7$, but appears to resemble Ho$_2$Ti$_2$O$_7$ more closely. The radial spin correlation function was also calculated for all three spin ices, showing distinguishable differences between Ho$_2$Ge$_2$O$_7$ and Ho$_2$Ti$_2$O$_7$ only in the third neighbor correlations, for which Ho$_2$Ge$_2$O$_7$ is somewhat less ferromagnetic.

The pyrochlore Ho$_2$Ge$_2$O$_7$ exhibits all the distinctive properties of a dipolar spin ice: a small, positive Curie-Weiss constant; Pauling zero-point entropy; magnetization which saturates to half the magnetic moment; a spin freezing transition in the ac susceptibility; and the characteristic magnetic diffuse scattering of spin ices. Reverse Monte Carlo refinements using powder diffuse scattering data suggest that the data are consistent with no violations of the ice rule at 50~mK, although powder data alone cannot exclude the possibility that a fraction of monopoles of up to 10\% remains present. The RMC refinements also allow us to predict the single crystal diffuse scattering for Ho$_2$Ge$_2$O$_7$, revealing the presence of spin correlations closely resembling the dipolar spin ice model. Thus, Ho$_2$Ge$_2$O$_7$ will be the best natural candidate for monopole studies involving neutron scattering.

\begin{acknowledgments}
A.M.H. is grateful to NSERC for funding and to A.~R.~Wildes for technical support at the Institut Laue-Langevin. H.J.S. acknowledges support from Vanier and MGS. C.R.W. acknowledges support through NSERC, CFI, and the ACS Petroleum Fund. J.A.M.P. and A.L.G. acknowledge STFC and EPSRC (No. EP/G004528/2) for funding. This work utilized facilities supported in part by the NSF through the Cooperative Agreement No. DMR-0654118 and the State of Florida. J.B.G. is grateful for financial support from NSF DMR-0904282, DMR-1122603 and the Robert A. Welch Foundation (No. F-1066). G. E. acknowledges funding by the Scientific User Facilities Division, Office of Basic Energy Sciences, U.S. Department of Energy. 
\end{acknowledgments}

\bibliography{Ho2Ge2O7_2}

\end{document}